# TIME VARIATIONS OF THE FORBUSH DECREASE DATA

[1]Koushik Ghosh and [2]Probhas Raychaudhuri

[1]Department of Mathematics
University Institute of technology
University of Burdwan
Burdwan-713 104
INDIA
Email:koushikg123@yahoo.co.uk

[2]Department of Applied Mathematics
University of Calcutta
92, A.P.C. Road, Kolkata-700 009
INDIA
Email:probhasprc@rediffmail.com

**ABSTRACT:**

We have used Simple Denoising Algorithm using Wavelet Transform on the daily Forbush decrease data from the year 1967 to 2003. For this data we observe periodicity around 5-6, 11, 13, 15 and 24 years. For all the obtained peaks corresponding confidence levels are higher than 95%. We observe that the periodicity of around 5-6 years is common to solar flare data, major proton event data and solar neutrino flux data. Because of that common periodicity, it is suggested that Forbush decrease with the solar flare data and major solar proton event data together with solar neutrino flux variations, behave similarly and may have a common origin.

**KEY WORDS:**

Simple Denoising Algorithm, Wavelet Transform, daily Forbush decrease data, periodicity.

**INTRODUCTION:**

A Forbush decrease is a rapid decrease in the observed galactic cosmic ray intensity following a coronal mass ejection (CME). It occurs due to the magnetic field of the plasma solar wind sweeping some of the galactic cosmic rays away from earth. The term Forbush decrease was named after the American physicist Scott E. Forbush, who studied cosmic rays in the 1930s and 1940s. The Forbush decrease is usually observed by particle detectors on Earth within a few days after the CME, and the decrease takes place over the following several days, the solar cosmic ray intensity returns to normal. Forbush decreases have also been observed by humans on Mir and the International Space Station (ISS) and by instruments onboard Pioneer 10 and 11 and Voyager 1 and 2, even past of the orbit of Neptune. The magnitude of a Forbush decrease depends on three factors:

(i) The size of the CME
(ii) The strength of the magnetic fields in the CME
(iii) The proximity of the CME to the Earth.

A Forbush decrease is sometimes defined as being a decrease of at least 10% of galactic cosmic rays on Earth, but ranges from about 3% to 20%. Reductions of 30% or more have been recorded a board the ISS. The overall rate of Forbush decrease tends to follow the 11-year sunspot cycle.

Applications of wavelets and multiresolution analysis to reaction engineering system from the point of view of process monitoring, fault detection, system analysis and so on, is an important topic and of current research interest [1]. The presence of noise in a time series data restricts one significantly to have meaningful information from the data. Noise in experimental data can also cause misleading conclusions [2]. Donoho and Johnstone [3] introduced two types of Denoising: linear denoising and nonlinear denoising. In linear denoising noise is assumed to be concentrated only on fine scales and that all the wavelet coefficients below these scales are cut off. Nonlinear denoising, on the other hand, treats noise reduction by either cutting off all coefficients below a certain threshold or reducing all coefficients by this threshold [1, 3]. The threshold values are obtained by statistical calculations and have been seen to depend on the standard deviation of the noise [4]. Ray et al [1] proposed a denoising algorithm making use of the discrete analogue of the Wavelet Transform (WT), which in many ways complements the well-known Fourier Transform (FT) procedure. The unique part of this proposal is that in this noise-reduction algorithm the threshold level for noise is identified automatically.

The detection of the peaks in a time series data does not completely serve our purpose. Further analysis is required to analyze the confidence levels of the obtained peaks to classify the strong peaks and weak peaks. We here modify the Integral Method in connection to this view using the idea developed by Azzini et al [5].

We have used this Simple Denoising Algorithm on the daily Forbush decrease data demonstrated in two series from the year 1967 to 2003 from Eroshenko et al [6]. Graphs are plotted to depict the periodic nature of the Forbush decrease data and the corresponding periodicities with levels of confidence are calculated.

**THEORY:**
**a) ALGORITHM BASED ON DISCRETE WAVELET TRANSFORM:**

The noise-reduction algorithm consists of the following steps [1]:
**Step 1:** In the first step, we differentiate the noisy signal x(t) to obtain the data $x_d(t)$, [1, 7] i.e.

$$x_d(t) = dx(t)/dt \qquad (1)$$

**Step 2:** We then take the Discrete Wavelet Transform of the data $x_d(t)$ and obtain wavelet coefficients $W_{j,k}$ at various dyadic scales j and displacements k. A dyadic scale is the scale whose numerical magnitude is equal to 2, rose to an integer exponent and is labelled by the exponent. Thus the dyadic scale j refers to a scale of size $2^j$. In other words, it indicates a resolution of $2^j$ data points. Thus a low value of j implies a finer resolution, while high j analyzes the signal at a larger resolution. This transform is the discrete analogue of continuous Wavelet Transform [1, 8] and it can be represented by the following formula

$$W_{j,k} = \int_{-\infty}^{\infty} x_d(t) \Psi_{j,k}(t) dt \qquad (2)$$

with

$$\Psi_{j,k}(t) = 2^{j/2} \Psi(2^j t - k)$$

where j, k are integers.

Taking the above formula (2) in a discrete sense and using (1) we have for n number of observations occurring at equal time intervals at t=1, 2, 3, …., n in which we have taken an appropriate time scale where the length of the equal time interval is taken to be unity:

$$W_{j,k} = x(n)\Psi_{j,k}(n) - x(1)\Psi_{j,k}(1) - \sum_{t=1}^{n} x(t) \Psi_{j,k}'(t) \qquad (3)$$

As for the wavelet function $\Psi(t)$, we have chosen Daubechies' compactly supported orthogonal function with four filter coefficients [9, 10].

**Step 3:** In this step we estimate the power $P_j$ [1] contained in different dyadic scales j, via

$$P_j(x) = \sum_k |W_{j,k}|^2 \qquad (j=1, 2, ….) \qquad (4)$$

By plotting the variation of $P_j$ with j, we see that it is possible to identify a scale $j_m$ at which the power due to noise falls of rapidly. This is important because it provides an automated means of detecting the threshold. Identification of the scale $j_m$ at which the power due to noise shows the first minimum allows us to reset all $W_{j,k}$ up to scale index $j_m$ to zero, that is $W_{j,k}=0$, for j=1, 2,…., $j_m$ [1].

**Step 4:** In the fourth step, we reconstruct the denoised data $\hat{x}_d(t)$ by taking the inverse transform of the coefficients $\hat{W}_{j,k}$

$$\hat{x}_d(t) = \sum_j \sum_k \hat{W}_{j,k} \Psi_{j,k}(t) \qquad (5)$$

This set of obtained $\hat{x}_d(t)$ gives measure of time variation in the signal. Upon differentiation the contribution due to white noise moves towards the finer scales because the process of differentiation converts the uncorrelated stochastic process to a first order moving average process and thereby distributes more energy to finer scales [1].

Finally we plot the graph of $\hat{x}_d(t)$ vs. t to obtain the corresponding peaks.

## b) INTEGRAL METHOD TO DETERMINE THE CONFIDENCE LEVELS OF PEAKS:

The obtained $\hat{x}_d$'s are processed for their corresponding confidence levels by this method [5]. First we make,

$$\sum_{i=1}^{m} |\hat{x}_d(T_i)| = A_1 \qquad (6)$$

where m is the total number of trial periods $T_i$ taken for analysis.
We take,

$$\Delta \hat{x}_d(T_i)/\Delta T_i = [\hat{x}_d(T_{i+1}) - \hat{x}_d(T_i)]/(T_{i+1} - T_i) = \Delta_1 \hat{x}_d(T_i) \qquad (7)$$

We take τ as the time $T_j$ where $\Delta_1 \hat{x}_d(T_i)$ is maximum. That means,

$$\tau = \text{argmax}[\hat{x}_d(T_i)] \qquad (8)$$

Next we consider

$$A_2 = \sum_{k=j-2}^{j+2} |\hat{x}_d(T_k)| \qquad (9)$$

Finally we set the ratio

$$S_0 = (A_2/A_1) \times 100 \qquad (10)$$

The value of $S_0$ gives the corresponding confidence level in percentage for the peak. The higher the score $S_0$ the higher is the probability to find a significant peak with respect to average signal amplitude.
In the first step we proceed by the above manner to determine the confidence level for the peak, where $\Delta_1 \hat{x}_d(T_i)$ is maximum.

Completing the analysis next we make a shift of $\hat{x}_d$ at the second highest value of $\Delta_1 \hat{x}_d(T_i)$ and proceed in the same way to determine the confidence level for the corresponding peak. Continuing the process we can find the confidence levels for all the possible peaks for our discussion.

## RESULTS:

For numerical simulation we take j=1, 2, ...., 5 and k=1, 2, ...., 5. Calculations yield different periodicities for the two series of daily data of Forbush effects [6] in the present analysis. An interesting observation for the presently considered algorithm is that the method has a tendency to locate the periods in comparatively larger time scale. So we search for the initial peaks of the Forbush decrease data by taking the initial time range as a whole and then restudy the case to obtain the initial peaks. Corresponding confidence levels are also calculated for the obtained peaks. The peaks obtained for the presently considered two series of data are presented in the tabular form with corresponding confidence levels given in brackets:

| Data | Periodicities (in years) |
|---|---|
| (1) First series of daily Forbush decrease data from 1967 to 2003 | 5.60 (99.89%), 11.34 (99.32%), 13.26 (97.27%), 14.39 (98.28%), 14.79 (96.48%), 15.10 (98.83%), 15.54 (98.17%), 15.94 (96.24%), 16.39 (99.31%), 24.45 (99.67%). |
| (2) Second series of daily Forbush decrease data from 1967 to 2003 | 5.60 (99.89%), 11.13 (99.64%), 13.26 (97.27%), 14.39 (98.28%), 14.79 (96.48%), 15.10 (98.83%), 15.54 (98.17%), 15.92 (96.42%), 24.45 (99.67%). |

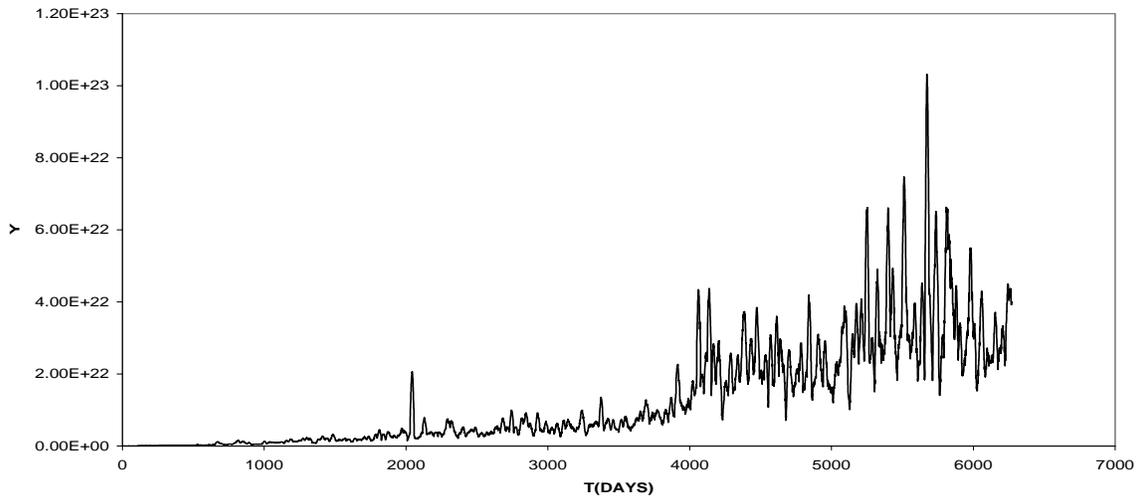

**FIG. 1: ANALYSIS OF DAILY FORBUSH DECREASE DATA(1ST TYPE) FROM 1ST JANUARY, 1967 TO 31ST DECEMBER, 2003 FOR INITIAL PEAKS BY SIMPLE DENOISING ALGORITHM USING WAVELET TRANSFORM**

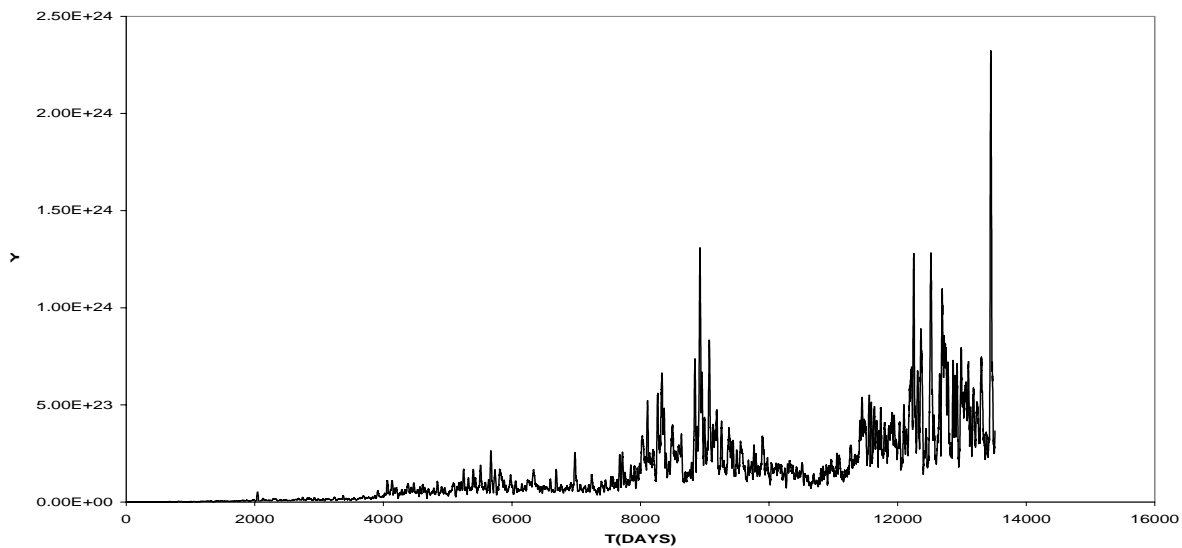

FIG. 2: ANALYSIS OF DAILY FORBUSH DECREASE DATA(1ST TYPE) FROM 1ST JANUARY, 1967 TO 31ST DECEMBER, 2003 BY SIMPLE DENOISING ALGORITHM USING WAVELET TRANSFORM

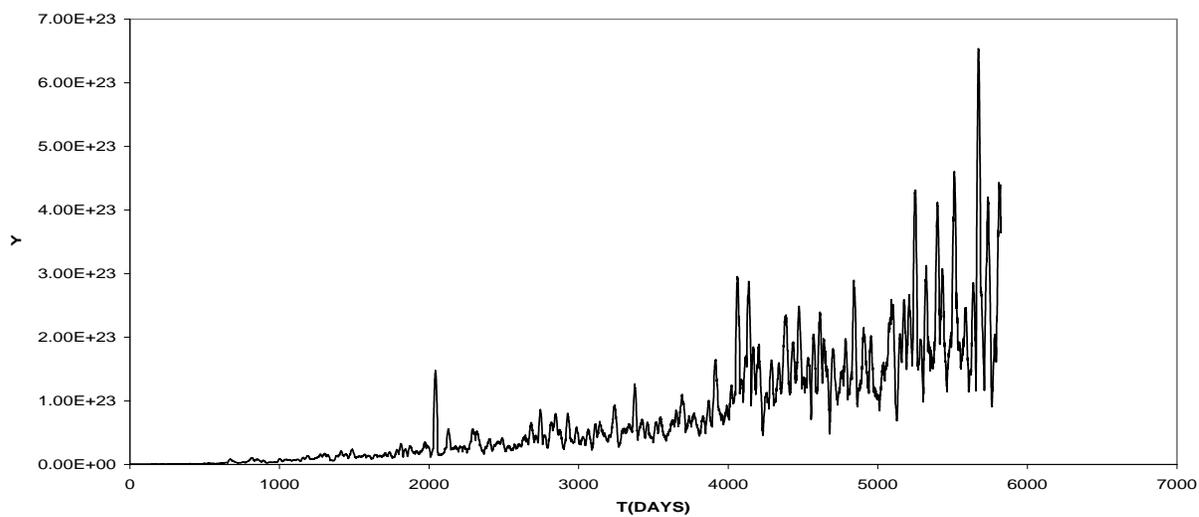

FIG. 3: ANALYSIS OF DAILY FORBUSH DECREASE DATA(2ND TYPE) FROM 1ST JANUARY, 1967 TO 31ST DECEMBER, 2003 FOR INITIAL PEAKS BY SIMPLE DENOISING ALGORITHM USING WAVELET TRANSFORM

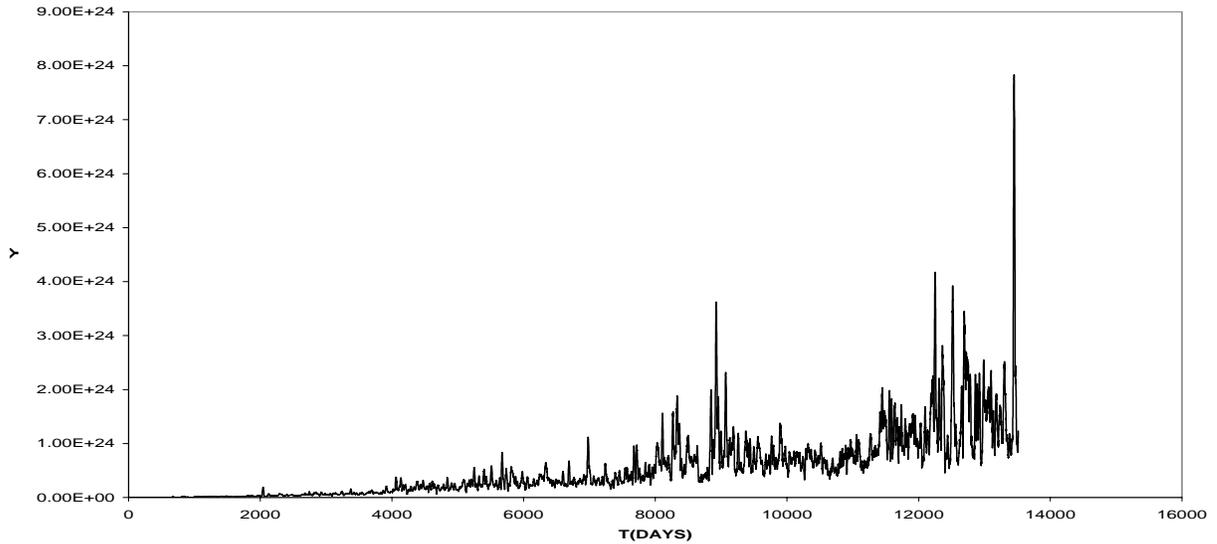

**FIG. 4: ANALYSIS OF DAILY FORBUSH DECREASE DATA(2ND TYPE) FROM 1ST JANUARY, 1967 TO 31ST DECEMBER, 2003 BY SIMPLE DENOISING ALGORITHM USING WAVELET TRANSFORM**

[In the above figures $\hat{X}_d(T)$ is taken to be Y]

## DISCUSSION:

Raychaudhuri [11] found that the Forbush decrease and solar flare data for the year 1976-1986 exhibit periods around 0.95, 2.4 and 4.75 years at >99% level of confidence while solar proton event data during the same time interval exhibit periods around 2.6 and 5.0 years at >95% level of confidence. The period of 5.60 (99.89% CL) years obtained here for both the series of Forbush decrease data is not appreciably different from 4.75 years found for Forbush decrease and solar flare data [11] and 5.0 years for solar proton event data [11]. Ghosh and Raychaudhuri [12] obtained almost similar period of 5.803 years for combined GALLEX-GNO solar neutrino flux data during May 1991 to December 2001. Thus the common periodicity of around 5 to 6 years in the solar neutrino flux data along with Forbush decrease data with corresponding solar flare data and major solar proton event data suggest that they are inter related and that a pulsating solar core may be their common origin [13, 14].

## REFERENCES:

[1] Roy, M., Kumar, V.R., Kulkarni, B.D., Sanderson, J., Rhodes, M. and Stappen, M.V., *AIChE Jour.*, **45 (11)**, 2461 (1999).
[2] Grassberger, P., Schreiber, T. and Schaffrath, C., *Int. Jour. Bifurcation Chaos*, **1**, 521 (1991).
[3] Donoho, D.L. and Johnstone, I.M., *J. Am. Stat. Assoc.*, **90**, 1200 (1995).
[4] Nason, G.P., Private Communication to Roy et al [1] (1994).

[5] Azzini, I., Dell' Anna, R., Ciocchetta, F., Demichelis, F., Sboner, A., Blanzieri, E. and Malossini, A., http://www.camda.duke.edu/camdo04/papers/days/thursday/azzini/paper.

[6] Belova, A.V., Buetikoferb, R., Eroshenko, E.A., Flueckigerb, E.O., Gushchinaa, R.T., Olenevaa, V.A. and Yankea, V.G., *Proceedings of 29$^{th}$ International Cosmic Ray Conference [Held in Pune, India, 2005]*, **1 (SH 2.6)**, 375 (2005).

[7] Constantinides, A., in *Applied Numerical Methods with personal computers*, McGraw-Hill, New York (1987).

[8] Holschneider, M., in *Wavelets: An Analysis Tool*, Clarendon Press, Oxford (1995).

[9] Daubechies, I., *Ten Lectures on Wavelets*, SIAM, Philadelphia (1990).

[10] Press, W.H., Flannery, B.P., Teukolsky, S.A. and Vetterling, W.T., in *Numerical Recipes*, Cambridge Univ. Press, Cambridge (1987).

[11] Raychaudhuri, P., *Solar Phys.*, **153**, 445 (1994).

[12] Ghosh, K. and Raychaudhuri, P., *Proceedings of the National Conference on Nonlinear Systems and Dynamics,NCNSD-2003(held in I.I.T.,Kharagpur,India,28-30 December,2003)*, 297 (2003).

[13] Raychaudhuri, P., *Astrophys. Space Sci.*, **13**, 231 (1971).

[14] Raychaudhuri, P., *Mod. Phys. Lett.*, **A8**, 1961 (1993).